\documentclass[10pt, a4paper, onecolumn]{article}

\usepackage[numbers]{natbib}

\usepackage[T1]{fontenc}
\usepackage{lmodern}

\usepackage{natbib}
\usepackage[utf8]{inputenc} 
\usepackage{booktabs}       
\usepackage{amsfonts}       
\usepackage{nicefrac}       
\usepackage{microtype}      
\usepackage{xcolor}         
\usepackage{graphicx}
\usepackage{amsmath,amssymb,amsfonts}
\usepackage{bm}

\usepackage{url}
\usepackage{hyperref}

\usepackage{algorithm}
\usepackage{algorithmicx}
\usepackage{algpseudocode}

\usepackage{array,multirow,graphicx}

\usepackage{geometry}
\newgeometry{vmargin={15mm}, hmargin={12mm,17mm}} 

\title{Resource saving taxonomy classification with k-mer distributions and machine learning}

\author{
	Wolfgang Fuhl
	\and
	Susanne Zabel
	\and
	Kay Nieselt
	}

\date{}

\begin{document}
	
	\maketitle
	
	\begin{abstract}
		Modern high throughput sequencing technologies like metagenomic sequencing generate millions of sequences which have to be classified based on their taxonomic rank. Modern approaches either apply local alignment and comparison to existing data sets like MMseqs2 or use deep neural networks as it is done in DeepMicrobes and BERTax. Alignment-based approaches are costly in terms of runtime, especially since databases get larger and larger. For the deep learning-based approaches, specialized hardware is necessary for a computation, which consumes large amounts of energy. In this paper, we propose to use $k$-mer distributions obtained from DNA as features to classify its taxonomic origin using machine learning approaches like the subspace $k$-nearest neighbors algorithm, neural networks or bagged decision trees. In addition, we propose a feature space data set balancing approach, which allows reducing the data set for training and improves the performance of the classifiers. By comparing performance, time, and memory consumption of our approach to those of state-of-the-art algorithms (BERTax and MMseqs2) using several datasets, we show that our approach improves the classification on the genus level and achieves comparable results for the superkingdom and phylum level. \\
		Link: \url{https://es-cloud.cs.uni-tuebingen.de/d/8e2ab8c3fdd444e1a135/?p=%2FTaxonomyClassification&mode=list}
	\end{abstract}

	\section{Introduction}
	Given a DNA sequence, one of the oldest problems in microbiology is to classify its microbial origin. This ranges from identifying the etiology of an infection from a blood sample of a patient to computing the bacterial composition of an environmental soil sample~\citep{jones2001scipy,pedersen2016human,somasekar2017viral,zhang2016viral}. Before the rise of genomic sequencing technologies, taxa identification comprised time-consuming sequential testing of potential candidates~\citep{pavia2011viral,venkatesan2013case}. Metagenomic sequencing enables the direct assignment of the taxonomic composition of a community. This facilitates rapid species detection and the discovery of novel species without relying on culture-dependent approaches~\citep{knights2011bayesian,loman2013culture}, and is, for example, widely applied in public health settings to improve and support diagnostics~\citep{chiu2019clinical,miller2013metagenomics}. 
	
	Metagenomic sequencing identifies nucleic acids in a sample. Those contain mixed populations of microorganisms, and the task is to assign them their reference genomes. This makes it possible to quantify the microbial composition. The ability to identify material from different kingdoms of organisms makes metagenomics an important and powerful tool. One of the primary challenges of the metagenomic analysis is to efficiently determine the species contained in the collected samples from the set of reads. This computational challenge is compounded by two important factors. Firstly, the use of high throughput sequencing technologies generates millions of sequences (of length 50–200 nt)~\citep{simon2019benchmarking}. Therefore, a large amount of reads has to be classified correctly in a reasonable amount of time and without wasting large amounts of energy. The most well-known tool for this task is BLAST (Basic Local Alignment and Search Tool). BLAST searches and aligns the collected samples against a database of genomic sequences to find the best fit~\citep{altschul1990basic}. While BLAST is one of the most sensitive tools for this task, it is computationally expensive and, therefore, not fast enough to be used for millions of reads generated by metagenomic sequencing. The second factor is the exponential increase in the number of sequenced microbial genomes in recent years. Therefore, the number of comparisons that need to be performed is constantly growing, which increases the runtime of BLAST or any tool which queries DNA/RNA sequences against databases. Another example is the number of reference genomes in databases like RefSeq (125,116 on December 5th, 2022) which possibly represent less than 6\% of all species~\citep{louca2019census} and this is an optimistic estimation~\citep{10.1371/journal.pbio.3001192,louca2021response}.
	
	Metagenomic analyses, therefore, require fast and resource saving algorithms which map the DNA sequences to taxonomic classes. In recent years, many tools have been proposed to accomplish the task of taxa classification given a DNA or RNA sequence. The most common technical approaches use local alignments, $k$-mers, Burrows–Wheeler transformations or hybrid methods~\citep{morgulis2008database,buchfink2015fast,ounit2015clark,bui2020cdkam,kim2016centrifuge,menzel2016fast,ainsworth2017k,steinegger2017mmseqs2,wood2019improved,brown2016sourmash,li2018minimap2}. Using local alignment-based approaches are usually very accurate, but relies on the database and have a large computational cost. In contrast, $k$-mer based approaches are usually very fast, but less accurate compared to the local alignment-based approaches~\citep{simon2019benchmarking}. More modern approaches using deep neural network (DNN) have also been applied to taxonomy classification. One of those approaches is DeepMicrobes~\citep{liang2020deepmicrobes}, which uses a $k$-mer (k = 12) as input feature representation and feeds it to a large neural network with long-short-term-memory cells. The recent published method BERTax~\citep{mock2022taxonomic} uses a transformer model from natural language processing (NLP), trained on large amounts of DNA sequences, to extract a feature representation of the DNA sequence~\citep{mock2022taxonomic}. This feature representation is then fed to a small, fully connected neural network for the final classification. Both approaches, BERTax and DeepMicrobes, show a comparable accuracy as the alignment-based methods and can be executed much faster with a GPU. For training, most large transformer models are trained multiple days on GPU clusters~\citep{lin2022language}. This limits the applicability of those approaches, since not every scientist has a computer with this specific hardware. In addition, the results of those methods are difficult to interpret and retraining of them has to be done by experts with the required hardware. In this work, we propose a simple feature based on $k$-mers. Instead of using $k$-mers directly, we compute the distribution over the $k$-mer occurrences in a DNA sequence as a summary statistic. This $k$-mer is given to a resource saving machine learning approach like a decision tree or subspace k-nearest neighbor classifiers for training, which are additionally more easy to interpret.
	
	In addition, we propose a data set reduction approach in the $k$-mer space, since DNA databases contain millions of entries. Using all entries for training consumes large amounts of memory and since all entries have to be processed during training, the computational load is also very high. Another disadvantage of using the entire database for training, is that the data is usually very unbalanced. This means that some taxonomies are overrepresented and in terms of the $k$-mer space are very dense at some locations in comparison to others.
	
	In summary, our contributions are:
	\begin{itemize}
		\item Using $k$-mer distributions with simple and resource saving machine learning methods for taxonomy classification.
		\item An data set balancing approach for $k$-mer features which reduces the amount of data for training.
	\end{itemize}

	\section{Related work}
	As of today more than 20 tools exist for metagenomic classification. The challenge today is to classify large amounts of reads from metagenomic sequencing as fast as possible~\citep{simon2019benchmarking}. Some of those tools require pre-computed databases against which the new sequences are matched. Modern approaches with deep neural networks (DNN) or other machine learning classifiers like random forests etc. contain this information in the learned parameters. In general there are two common approaches, the first is taxonomic binning in which each sequence read receives a class from the algorithm. The second approach is taxonomic profiling, where the relative abundances of taxa within a data set is reported. For a conversion from the binning results to a taxa profile, it is only necessary to summing up the individual classifications~\citep{simon2019benchmarking}.
	
	One approach to speed up the classification for large amounts of reads in alignment-based methods is to reduce the number of candidate hits. Therefore, the first step searches for perfect sequence matches in the reference sequences or using full-text index in minute space~\citep{ferragina2000opportunistic}. While this approach speeds up the classification, it is not as sensitive as BLAST.
	
	All tools can be divided into four categories, DNA-to-DNA classification, DNA-to-protein classification, marker-based classification, and the direct classification with a DNN for example. The first three approaches rely on comparing the reads against a database. Examples are: Kraken2~\citep{wood2019improved}, sourmash~\citep{brown2016sourmash}, MMseqs2~\citep{steinegger2017mmseqs2}, or minimap2~\citep{li2018minimap2}. The fourth approach is based on modern machine learning algorithms (Like DeepMicrobes~\citep{liang2020deepmicrobes}, BERTax~\citep{mock2022taxonomic}, Convolutional Neural Network - Relative Abundance Index (CNN-RAI)~\citep{karagoz2021taxonomic}, or GeNet~\citep{rojas2019genet}). DNA-to-protein tools are computationally more expensive than DNA-to-DNA tools since they have to search through all possible DNA-to-amino acid translations, but they are more sensitive~\citep{altschul1990basic}. Since DNA-to-protein tools target only coding sequences, they are not applicable to classify non-coding sequencing reads.
	
	For marker-based classifiers, the database consists only of a subset of gene sequences instead of whole genomes. The highly conserved 16S rRNA sequence for example is one of those widely used single marker genes for bacterial metagenomics~\citep{edgar2018updating,yarza2014uniting}. MetaPhlAn2 is one of the representatives of marker-based classifiers. Rather than using just one marker gene, it uses about one million marker genes  from different gene families for taxonomic classification~\citep{truong2015metaphlan2}. While the usage of only a subset of gene sequences makes those methods very resource saving and efficient, the marker sequences can introduce a bias when they are not evenly distributed~\citep{d2016comprehensive}.
	
	One of the current state-of-the-art representative of the database methods is Kraken2~\citep{wood2019improved}. This method uses $k$-mers, the lowest common ancestor (LCA), and hashes to compare sequences in the reference database against the new sequences. The novel part of Kraken 2 in contrast to Kraken is the use of the hash tables for faster indexing and reduction of the memory requirements. Sourmash~\citep{brown2016sourmash} uses MinHash sketches computed on DNA sequences as reference. This reduces the size of the database and the comparison of hashes can be done fast and evaluated via the Jaccard index. MMseqs2~\citep{steinegger2017mmseqs2} uses $k$-mers for a fast local matching in the first stage. Afterwards, an ungapped local alignment is computed, followed by a gapped alignment. Overall this is a costly approach and delivers good results for speed optimization, a lot of work has gone into the parallelization of the software. In minimap2~\citep{li2018minimap2} uses split alignments and employs concave gap costs for large insertions or deletions. Additionally, a new heuristic was proposed to reduce the amount of spurious alignments. Overall, minmap2 has been shown to be 3–4 times faster compared to mainstream short-read mappers.
	
	With the advances in machine learning from recent years, there are also numerous direct approaches in the literature. DeepMicrobes~\citep{liang2020deepmicrobes} is a deep learning architecture consisting of many LSTM (Long short term memory) cells followed by self attention and fully connected layers for the final classification. DeepMicrobes utilizes a $k$-mer embedding as the first layer of the DNN. GeNet~\citep{rojas2019genet} is a convolutional neural network with the ResNet architecture. The input to GeNet are the indexes encoded by the 4 integers 0,1,2,3 of the four nucleotides. The authors of Convolutional Neural Network - Relative Abundance Index (CNN-RAI)~\citep{karagoz2021taxonomic} also use a convolutional neural network. The input to their model are $k$-mers and the relative abundance index score. Seq2Species~\citep{busia2019deep} uses a deep learning architecture consisting of three depthwise separable convolutional layers and followed by two or three fully connected stages. The input to their DNN is a four-dimensional vector per amino acid, and for ambiguity codes a probability distribution of those for dimensions is computed. MetagenomicDC~\citep{fiannaca2018deep} uses a deep neural network with $k$-mers as input (k=5). As DNN they analyzed convolutional neural networks and deep belief network. Seq2Species and MetagenomicDC are both restricted to 16S sequences, which limits the applicability of those approaches. CHEER~\citep{shang2021cheer} is an inception architecture with convolution and fully connected stages. The main purpose of CHEER is RNA virus taxonomy classification. The entire architecture consists of multiple CNNs in a hierarchy, and the input can be either a four-dimensional vector per amino acid or a skip-gram based word embedding. Another deep learning architecture for virus classification was proposed in~\citep{fernandes2021novel}. They use a convolutional neural network with many drop out stages on four-dimensional vector representations per amino acid. The novel part of their approach is the pruning and weight quantization during training.
	
	The difference of our approach to the state of the art is that we use $k$-mer distributionscombined with simple machine learning methods. Additionally, we propose a data set reduction method, which balances the data set based on the $k$-mer feature space.

	\section{Method}
	\begin{figure}[!tb]
		\centering
		\includegraphics[width=0.4\textwidth]{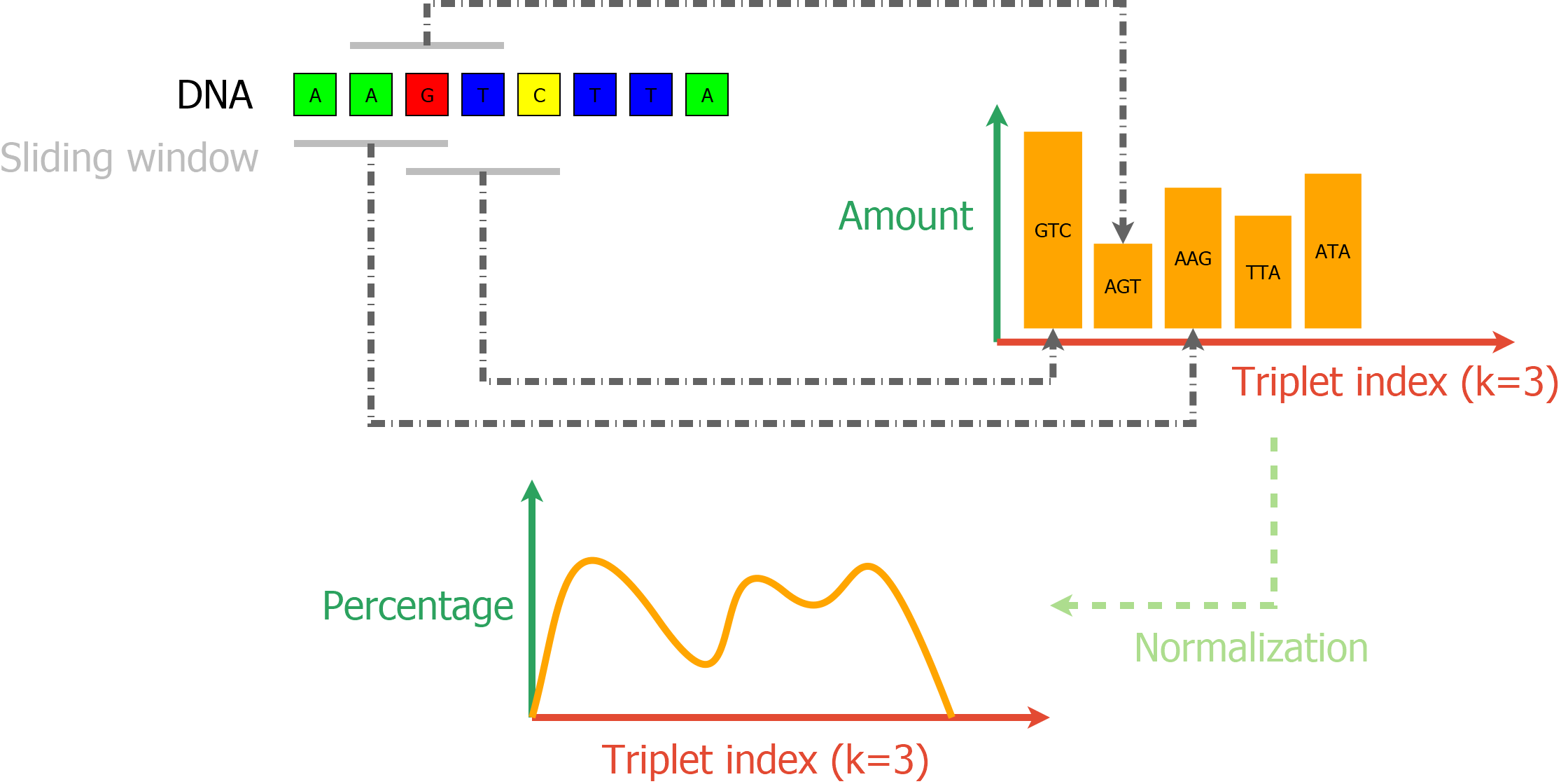}
		\caption{The feature definition process. At first, a histogram over all occurring $k$-mers (here $k=3$) is computed using a sliding window. Afterterwards, normalization converts the abundance values into relative frequencies according to  Equation~\ref{eq:featureNorm}. The resulting relative $k$-mer distribution is used as a feature for machine learning methods.}
		\label{fig:feature}
	\end{figure}

	The computation of our $k$-mer distribution is shown in Figure~\ref{fig:feature} and Algorithm~\ref{al:kmergeneral}. First, we compute the occurrence of all $k$-mers in a given DNA sequence using a sliding window. We decided to use a sliding window, since the given DNA sequences are not aligned. 
	
	\begin{equation}
		kmer_i=\frac{kmer_i}{\sum_{j=1}^{n} kmer_j}
		\label{eq:featureNorm}
	\end{equation}
	
	After the histogram is computed, we normalize it, resulting in a relative $k$-mer distribution, where the relative frequency of each $k$-mer $kmer_i$ is defined by Equation~\ref{eq:featureNorm}.

	
	\begin{algorithm}
		\caption{The general algorithm to compute the $k$-mer distribution for any $k$. We used the indices 1, 2, 3, 4 for the nucleotides A, G, T, C. Any letter in the read which is not equal to either of the 4 nucleotides, is marked as invalid and ignored. This algorithm is based on array indexing which starts with 1.}
		\label{al:kmergeneral}
		\begin{algorithmic}[1]
			\Require $k, Data_{seq}$
			\State $kmer=zeros(4^k)$
			\For{$d_i=1:size(Data_{seq})-(k-1)$}
			\State $valid\_sequence=1$
			\State $index=1$
			\State $sum=0$
			\For{$k_i=1:k$}
			\If{$Data(d_i + (k_i-1)) \in \{1-4\}$} //A,G,T,C
			\State $index=index + (Data_{seq}(d_i + (k_i-1)) - 1) *(4^{k_i-1})$
			\Else
			\State $valid\_sequence=0$
			\State $break~loop$
			\EndIf
			\EndFor
			\If{$valid\_sequence==1$}
			\State $kmer(index)=kmer(index)+1$
			\State $sum=sum+1$
			\EndIf
			\EndFor
			\State $kmer=\frac{kmer}{sum}$ //Normalize the k-mer histogram
			\State $return~kmer$
		\end{algorithmic}
	\end{algorithm}
	
	
	\begin{algorithm}
		\caption{The data balancing algorithm based on the $k$-mer feature space. This algorithm is a fast approach to improve the $k$-mer feature space balancing. It only considers the maximal occurrence of a feature space index for balancing (Line 7). In Line 6 the algorithm checks whether a feature space index was already found or not. This can be computed very fast with a hashing function, since otherwise a comparison to all indexes would be necessary.}
		\label{al:datablalance}
		\begin{algorithmic}[1]
			\Require $MaxOutputSize, Data_{k-mer}, GridSize$
			\State $NewData=\emptyset$
			\State $GridCounter=\emptyset$
			\While{$size(NewData) < MaxOutputSize$}
			\State $X = RandomFrom(Data_{k-mer})$	//Select a random $k$-mer distribution
			\State $XGrid = round(X*GridSize)$ //Grid index for $k$-mer distribution
			\If{$XGrid \in GridCounter$}
			\If{$GridCounter(XGrid)<Max(GridCounter)$}
			\State $Add(NewData,X)$
			\State $GridCounter(XGrid) += 1$
			\EndIf
			\Else
			\State $Add(NewData,X)$
			\State $Add(GridCounter,XGrid)$
			\State $GridCounter(XGrid) = 1$
			\EndIf
			\EndWhile
			\State $return~NewData$
		\end{algorithmic}
	\end{algorithm}
	
	Since the source of samples for DNA sequencing experiments is highly diverse and the collection of DNA sequences is usually driven by certain needs or specific research interests, we expect the data to be unbalanced. A class-based balancing is not sufficient in our setting, since despite of having a balanced set of classes the features behind it might be identical or too similar. In addition, we want to reduce the amount of data for training to improve time and memory consumption. The feature abundance itself also has an impact on the machine learning model. In deep neural networks, for example, it influences the gradient direction as high abundant features contribute more than low abundant ones. For other approaches like decision trees it influences split criteria like the GINI index or entropy and therefore, decision borders are misplaced. Hence, we decided to balance the data set based on the feature space. Since the feature space consists of probabilities which are floating-point numbers, we used a binning or grid-based approach. This means that each feature dimension is divided into a set of bins or that a one dimensional grid is overlaid. Here $X$ is the $k$-mer distribution and each bin is multiplied by the grid size. Afterwards, each bin is rounded, and we get integer steps per feature dimension. 
	The Algorithm~\ref{al:datablalance} starts by randomly selecting a DNA sequence (Line 4) and computing the grid index (Line 5). Afterwards, it checks whether this grid index is already in the samples or not. If not, the sequence is added to the new data set, otherwise it has to be lower than the current maximum of all stored grid indexes. The algorithm stops when the desired size of the data set is reached, which can be specified by $MaxOutputSize$. In our experiments, we added 30\% from $MaxOutputSize$ randomly to speed up the balancing approach. In addition, we ensured that those 30\% contain all classes from the taxonomy level. Since it is possible, that there is no new entry to the $GridCounter$ in the data set we integrated an additional check, which adds the current sample after 10 consecutive tries without a new insertion.
	
	\begin{figure}[!tb]
		\centering
		\includegraphics[width=0.48\textwidth]{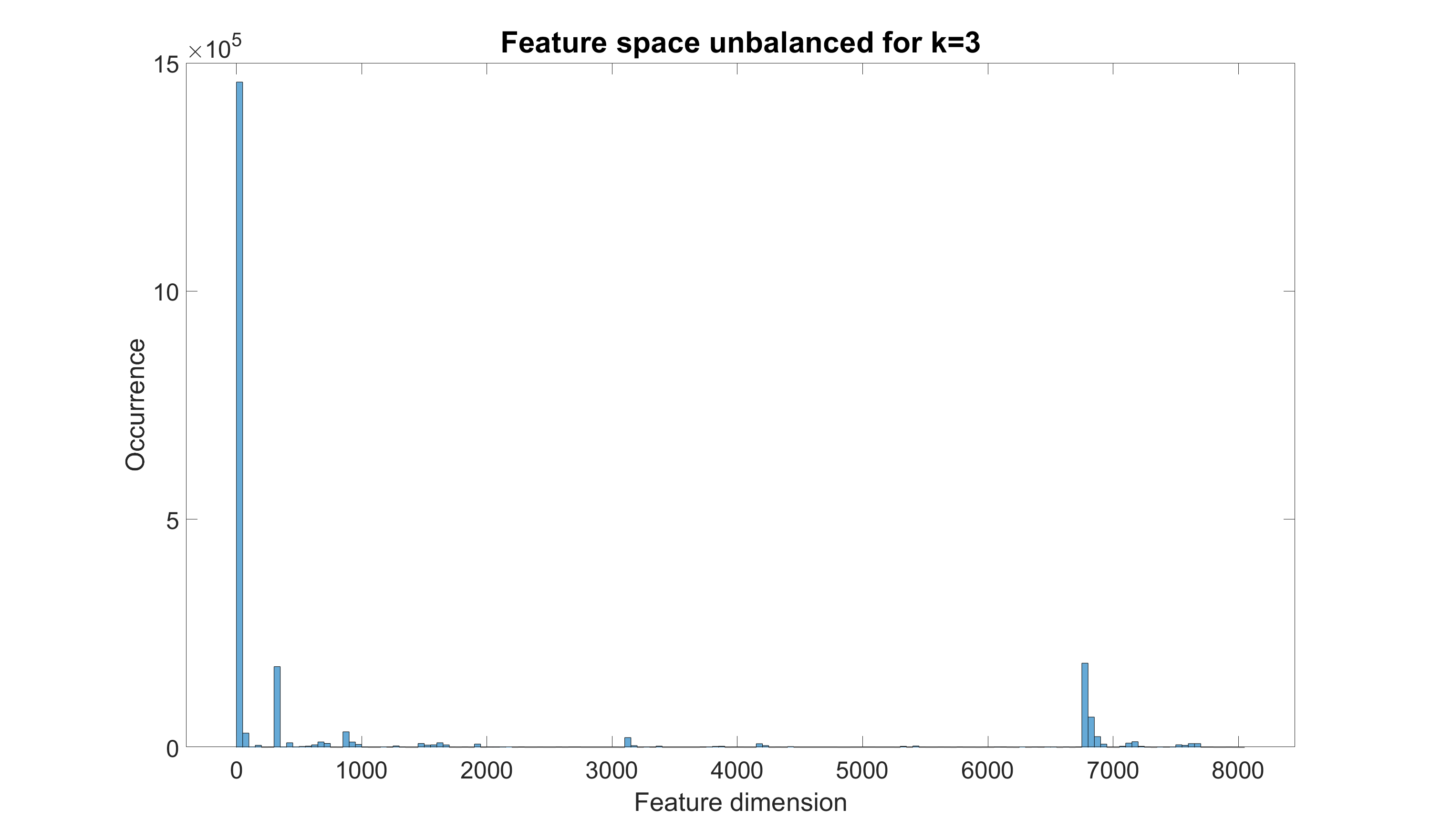}
		\includegraphics[width=0.48\textwidth]{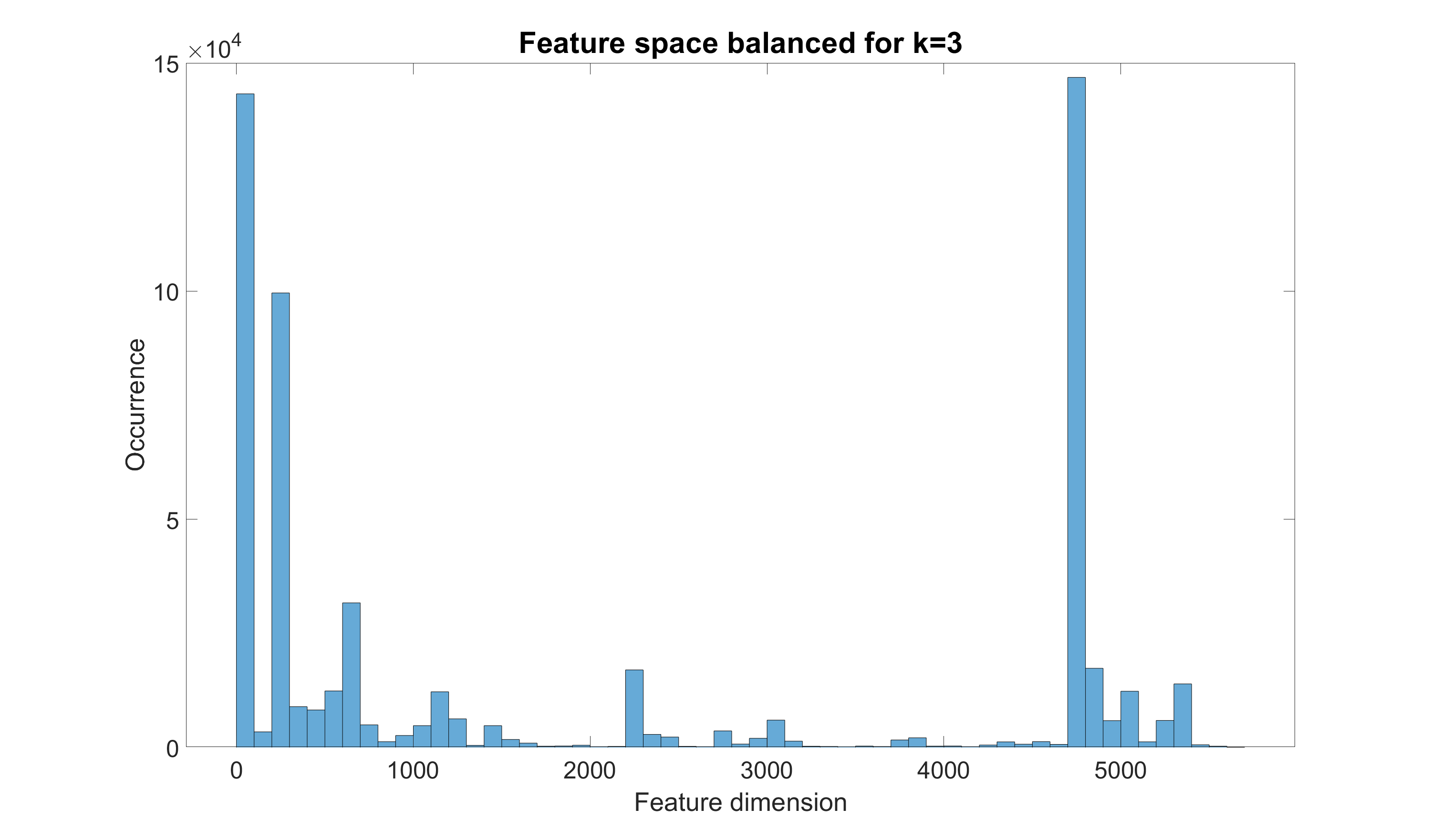}
		\caption{The results of the feature space balancing with Algorithm~\ref{al:datablalance} for $k=3$ on the non-similar data set. In the top histogram, the feature space occurrences for the 3-mer is shown using all data and a $GridSize$ of 10. Note that the y-axis is given in $10^5$. The bottom plot shows the feature space occurrences after  Algorithm~\ref{al:datablalance}. Since we only balance it based on the maximum, it is not uniformly distributed and a uniform distribution would mean to discard too many data samples. In the bottom plot, the y-axis is now given in $10^4$.}
		\label{fig:balance}
	\end{figure}
	
	Figure~\ref{fig:balance} shows the non-similar data set from~\citep{mock2022taxonomic} in the $k$-mer feature space for $k=3$ and a grid size of 10 per histogram bin. As can be seen in the top histogram, the data set is heavily unbalanced in the feature space. This means that some bacteria or viruses are more often sequenced compared to others, a characteristic which is very typical for metagenomics data sets. After applying Algorithm~\ref{al:datablalance}, using the maximal occurrence of the feature histogram, the data set appears much more balanced. In general, it would be even better to uniformly balance the data set, but this is not possible since there is no representative for large parts of the possible indexes in the feature space (All distributions which do not sum to 1). An alternative would be to balance all the available features, but this would mean, that for each bin only the lowest count of occurrences is allowed, or we would need to copy/augment all lower occurrences until we reach the maximum. With those approaches, we would either discard too much data or increase the data set, so it will require large amounts of resources during training. Our approach (Algorithm~\ref{al:datablalance}) aims to compromise both alternative approaches. First we reduce the data set and second we balance it in the feature space for the occurrences.
	
	\section{Evaluation}
	For our evaluations, we used two of the four data sets provided by ~\citep{mock2022taxonomic}. The four data sets contain sequences of length 1500 nt and are called pretraining, similar, non-similar, and final in this work corresponding to pretraining, closely related, distantly related, and final from ~\citep{mock2022taxonomic}. We downloaded the data sets on the 10th of September via \href{https://osf.io/qg6mv/}{https://osf.io/qg6mv/}. Since the similar data set is not challenging and the pretraining data set contains 100\% of the non-similar test data set as well as 46.89\% of the final test data set, we only used the non-similar and final data set with the given splits for our evaluations.
	
	For the non-similar (distantly) related data set, the authors tried to select sequences without (taxonomically) closely relationships. This was achieved by the authors with a genus separation between training and test sets. This means, that no sample in the test set exist in the training set. As target taxonomy levels, they used phylum and superkingdom. For phylum 30 classes are in the data set and for superkingdom 4. The test set consists of 53,400 and the training set of 2,245,416 sequences. For the training and validation split, they used 95\% of the training set for training and the remaining 5\% for validation. This split was done randomly.
	
	The last data set which is called final was created to be even more challenging. Here the authors provided 5,311,920 for training and 88,000 for testing. In this data set the taxonomy levels are superkingdom, phylum, and genus. For superkingdom there are 4 classes in the data set, for phylum there are 43 classes as well as the unknown class, and for genus there are 155 classes as well as the unknown class. For more details about the data set, we refer to the BERTax paper~\citep{mock2022taxonomic}.
	
	All our results are given in the macro average precision (MAP) metric as specified in Equation~\ref{eq:macroavgperc}.
	\begin{equation}
		MAP = \frac{\sum_{i=1}^{Classes} \frac{TP_i}{TP_i + FP_i}}{Classes}
		\label{eq:macroavgperc}
	\end{equation}
	Here,  $TP_i$ are the true positives for class $i$, $FP_i$ are the false positives for class $i$, and $Classes$ is the total number of classes.
	
	For the evaluation and training, we used three different servers. Two with an AMD Ryzen 9 3950X 16-Core (32 Threads) Processor (3.50 GHz) and 64 GB DDR4 memory. The third one has two Intel Xeon E5-2623 v3 CPUs (3.00 GHz) and 256 GB ECC DDR4 memory. For the runtime and resource consumption evaluation, we used only the first server with an AMD Ryzen 9 3950X CPU. The software we used is Matlab 2022b. In addition, we did not use any data augmentation like BERTax or DeepMicrobes did, which makes our results easy to reproduce. For the data set reduction, we evaluated different sizes and used the best performing size in terms of MAP for the non-similar and final data set. In both cases, it was approximately one third of the training set size. This means we reduced the training set amount for the non-similar data set to 600,000 and for the final data set to 1,600,000.
	
	For the classification, we used different machine learning models as ensemble. The first method is the subspace K nearest neighbor classifier. The method works the same way as the K nearest neighbor classifier (KNN) but only uses a subset of the available features. We always set $K=2$ and the subset of features to $\frac{feature~count}{2} + \sqrt(feature~count)$. The bagged decision tree classifier computes statistically optimal features and thresholds in an iterative manner. For this usually entropy, information gain, residence, etc. are used. In contrast to boosted trees, bagged trees work according to the major voting principle. This means that in the end, the ensemble selects the class with the most votes after the execution of all trees. 
	Other machine learning approaches like the support vector machine (SVM) and discriminant analysis were also trained as ensembles but will not be described in more detail. For comparison we also trained a small neural network with one hidden layer of size 512 and the ReLu activation function. 
	
	\begin{table*}
		\centering
		\caption{Evaluation of different grid sizes as well as different $k$-mer and the comparison to the usage of all data samples compared to the reduced data set. For all evaluations, we used the bagged decision trees. The reduced data sets, indicated with a specified grid size on top, used an ensemble size of 500 ecpet the $k$-mer 4 and 5 where we only could use 200 due to memory limitations. For the all data samples case, we only used an ensemble size of 200 due to memory limitations. We tested an ensemble size of 500 for the all data samples case only for the 3-mer in the non-similar data set and achieved 56.73 MAP which already outperforms the state-of-the-art results from Table~\ref{tbl:sotaeval}. The best results are reported in bold.}
		\label{tbl:grideval}
		\begin{tabular}{cccccccc}
			Data set & $k$-mer & All data & $GridSize=10$ & $GridSize=20$ & $GridSize=30$ & $GridSize=50$ & $GridSize=100$ \\ \hline
			\multirow{3}{*}{non-similar} & 1  & 0.1267 & 0.2834 & 0.2840 & 0.2843 & 0.2849 & 0.2835 \\
			& 2  & 0.4362 & 0.5829 & 0.5783 & 0.5816 & 0.5835 & 0.5828 \\
			\multirow{1}{*}{Phylum}& 3  & 0.5397 & \textbf{0.6971} & 0.6910 & 0.6928 & 0.6964 & 0.6936 \\ 
			& 4  & 0.4982 & 0.6562 & 0.6282 & 0.6582 & 0.6605 & 0.6452 \\ 
			& 5  & 0.3242 & 0.5728 & 0.6036 & 0.6134 & 0.5708 & 0.5372 \\ \hline
			\multirow{3}{*}{non-similar} & 1  & 0.3718 & 0.7658 & 0.7661 & 0.7671 & 0.7668 & 0.7667 \\
			& 2  & 0.6774 & 0.7916 & 0.7909 & 0.7925 & 0.7914 & 0.7902 \\
			\multirow{1}{*}{Superkingdom}& 3  & 0.7433 & 0.8164 & \textbf{0.8171} & 0.8153 & 0.8149 & 0.8160\\
			& 4  & 0.7252 & 0.7911 & 0.7881 & 0.7896 & 0.7912 & 0.7877\\
			& 5  & 0.6496 & 0.7693 & 0.7069 & 0.7439 & 0.7478 & 0.7693\\
		\end{tabular}
	\end{table*}
	
	Table~\ref{tbl:grideval} shows the MAP results on the non-similar data set. We evaluated the training on the entire samples versus the reduced set with different grid sizes. As can be seen in Table~\ref{tbl:grideval}, the grid size itself has a lower impact on the overall MAP. The variations in the results are mainly from the random selection of the balancing approach. Overall, the reduced set significantly improves the results and also allows us to use larger ensembles, which also improve the MAP. We could not train everything with an ensemble size of 500 due to the memory limitations of our servers. For the all data case, where we used the entire training set, it can be seen that we already achieve similar results as the state-of-the-art methods (Comparison to Table~\ref{tbl:sotaeval}). For the 3-mer distribution on the phylum taxonomy level we trained the ensemble once with a size of 500 and achieved 56.73 MAP which outperforms the state-of-the-art without combinations of multiple algorithms. In case of the 4-mer and 5-mer, we also had to use lower ensemble sizes, but overall there are outperformed by the 3-mer. This is of course due to the fixed size of 1500 nucleotides of the data set, which leads to sparse 4-mer and 5-mer representations. For larger sequences, those representations could perform much better. The reverse case where only smaller sequence sizes are available could benefit the smaller $k$-mers, but this is out of the scope of this work.
	
	\begin{table}
		\centering
		\caption{Comparison of concatenated $k$-mers based on MAP with the single usage on the reduced non-similar data set computed with grid size 10. We used bagged decision trees for all evaluations, with an ensemble size of 500 for the $k$-mers 1 to 3 and an ensemble size of 200 for the $k$-mer sizes 4 and 5. For the concatenations, we used an ensemble size of 200 due to memory limitations. The best results are reported in bold.}
		\label{tbl:combieeval}
		\begin{tabular}{ccc}
			$k$-mer  & non-similar phylum & non-similar superkingdom\\ \hline
			1  & 0.2834 & 0.7658 \\
			2  & 0.5829 & 0.7916 \\
			3  & \textbf{0.6971} & \textbf{0.8164} \\
			4  & 0.6562 & 0.7911 \\
			5  & 0.5728 & 0.7693 \\
			2+3  & 0.6763 & 0.8150 \\
			2+4  & 0.6498 & 0.7983 \\
			3+4  & 0.6475 & 0.8016 \\
		\end{tabular}
	\end{table}
	
	In Table~\ref{tbl:combieeval} we evaluated different concatenations of the $k$-mer distributions and compared them to the separate usage of each $k$-mer alone. As can be seen in Table~\ref{tbl:combieeval} the 3-mer still outperforms the other approaches, but we had to train the concatenations again with an ensemble size of 200 whereas the 3-mer is trained with an ensemble size of 500. This means that especially the 2-mer concatenated with the 3-mer is a good candidate for slightly better results than the 3-mer alone. Overall, none of the concatenations lead to a large improvement on the reduced non-similar data set, and we only report here a fraction of the combinatorial results. In general the concatenation could lead to large improvements especially if the sequences vary in size which is usually the case for metagenomic sequencing but since here also smaller sequences are produced, smaller $k$-mers could be a better choice, too.
	
	\begin{table}
		\centering
		\caption{Comparison of different machine learning approaches on the reduced non-similar data set computed with grid size 10 and the 3-mer distribution. The metric is MAP. For the bagged decision trees we used an ensemble size of 500 and for the subspace KNN the parameters specified in the beginning of the evaluation section. All other approaches use the standard parameters of Matlab and are trained as ensembles or error correction code classifiers. The best results are reported in bold.}
		\label{tbl:mleval}
		\begin{tabular}{lcc}
			& \multicolumn{2}{c}{non-similar} \\
			ML Method  & phylum &  superkingdom \\ \hline
			Bagged Trees  & \textbf{0.6971} & 0.8164 \\
			Subspace KNN & 0.6756 & \textbf{0.8823} \\
			Subspace Discriminant & 0.4894 & 0.6020 \\
			Linear SVM & 0.4505 & 0.6462 \\
			Gaussian SVM & 0.5705 & 0.6635 \\
			Polynomial SVM & 0.6013 & 0.6737 \\
			Neural Network 512 & 0.6598 & 0.8121 \\
		\end{tabular}
	\end{table}
	
	Table~\ref{tbl:mleval} shows a comparison of different machine learning approaches with the 3-mer distribution on the reduced non-similar data set computed with grid size 10. All the used machine learning approaches are trained as ensembles or in the case of the support vector machine as binary error correction code classifiers. For the phylum level, the bagged decision trees outperform the other approaches and for the superkingdom level, the subspace KNN performed best. Based on the taxonomy level, the 4-mer also outperforms the 3-mer, which is shown in Table~\ref{tbl:sotaeval} where we compare our approaches to the state-of-the-art. Overall, the best approaches in our evaluations are the bagged decision trees and the subspace KNN. We also evaluated small neural networks with one hidden layer which performed similar to the bagged decision trees and the subspace KNN, but were not as good in terms of MAP and require a validation set to stop the training. Deep convolutional neural networks or transformer models are also applicable to our feature, but since those require special hardware (GPU) and have a higher energy consumption, we omitted those evaluations.
	
	\begin{table*}
		\centering
		\caption{Comparison of our approach based on MAP to the state-of-the-art. We used the state-of-the-art results from the BERTax paper to avoid unnecessary energy waste. Different combinations of the proposed approach are shown. For the subspace KNN we used the parameters as described in the beginning of the evaluation. For the bagged decision trees, we used an ensemble size of 500 for the non-similar data set and an ensemble size of 200 for the final data set due to memory limitations. For all evaluations of the proposed approach, we used the reduced data set computed with grid size 10. It also has to be noted, that our approach could be combined too with MMseqs2 as well as BERTax or DeepMicrobes. The best results are reported in bold.}
		\label{tbl:sotaeval}
		\begin{tabular}{llccccc}
			&  & \multicolumn{2}{c}{non-similar} & \multicolumn{3}{c}{final} \\
			& Method & superkingdom & phylum & superkingdom & phylum & genus \\ \hline
			\multirow{7}{*}{State-of-the-art}& MMseqs2 & 0.6276 & 0.4136 & 0.9694 & 0.9290 & 0.7476 \\
			& MMseqs2 tax. & 0.6747 & 0.4344 & 0.9811 & 0.9347 & 0.7509 \\
			& minimap2 & 0.4412 & 0.2003 & 0.9346 & 0.8671 & 0.6668 \\
			& Kraken2 & 0.4436 & 0.1950 & 0.9365 & 0.8713 & 0.7056 \\
			& sourmash & 0.2514 & 0.0348 & 0.3104 & 0.0800 & 0.0307 \\
			& DeepMicrobes & 0.6725 & 0.3661 & 0.9813 & 0.9211 & 0.6643 \\
			& BERTax & \textbf{0.9006} & 0.5410 & 0.9862 & 0.9510 & 0.6692 \\ \hline
			\multirow{2}{*}{Combinations} & MMseqs2+DeepMicrobes & 0.7691 & 0.5051 & 0.9723 & 0.9320 & 0.7785 \\
			& MMseqs2+BERTax & 0.8757 & 0.6034 & \textbf{0.9976} & \textbf{0.9783} & 0.7933 \\ \hline
			\multirow{4}{*}{Proposed}  &  subspace KNN 3-mer & 0.8823 & 0.6756 & 0.9828 & 0.9316 & 0.8108 \\ 
			&  subspace KNN 4-mer & 0.8803 & 0.6577 & 0.9907 & 0.9553 & \textbf{0.8643} \\
			&  Bagged Trees 3-mer & 0.8164 & \textbf{0.6971} & 0.9251 & 0.8517 & 0.7610 \\
			&  Bagged Trees 4-mer& 0.7911 & 0.6562 & 0.9198 & 0.8513 & 0.6911 \\
			&  Neural Network 512 3-mer & 0.8121 & 0.6598 & 0.8763 & 0.8530 & 0.5770 \\
			&  Neural Network 512 4-mer & 0.8242 & 0.6733 & 0.9392 & 0.8985 & 0.6285 \\ \hline
			& Chance level & 0.25 & 0.0333 & 0.25 & 0.0227 & 0.0064 \\
		\end{tabular}
	\end{table*}
	
	Table~\ref{tbl:sotaeval} shows the results of our approach with different configurations to the state-of-the-art algorithms. For the comparison of the algorithms alone, BERTax achieves the best result for the superkingdom taxonomy level on the non-similar data set. In all other evaluations, our approach outperforms the state-of-the-art results. The combinations of MMseqs2 with BERTax achieve slightly better results for the superkingdom and phylum taxonomy levels on the final data set. Here it has to be mentioned, that our approach could be combined with the other algorithms too and our $k$-mer distribution and balancing could also be integrated into the MMseqs2 algorithm. A disadvantage of our approach, is that not one configuration leads to the best result. The bagged decision trees seem to be the best method for the phylum level taxonomy classification in the non-similar data set, and the subspace KNN with the 4-mer distribution are the best for the final data set. 
	
	
	
	\begin{table}
		\centering
		\caption{The runtime on a single CPU core per sequence and memory consumption comparison results. For all approaches, we measured the execution time per sample and the total memory consumption of the tool. The bagged decision trees use an ensemble size of 500. For the training of the bagged decision trees and the subspace KNN, we used the reduced set of the non-similar data set computed with grid size 10 for the 3-mer. NN-512 is a neural network with one hidden layer consisting of 512 neurons. \textit{(*) For MMseqs2, we used the provided executable and extracted the total run time information from the output. It also used all 32 threads of the CPU. (**) Converted to C++ functions.}}
		\label{tbl:timeeval}
		\begin{tabular}{lcc}
			Method & Memory  & Execution time \\
			& consumption (GB) & per sequence (Sec.) \\ \hline
			MMseqs2 (*) & 30 & 208.5 \\
			BERTax & 0.5 & 3.6243 \\
			subspace KNN & 16 & 1.4315 \\
			Bagged Trees (**) & 0.7 & 0.261 \\
			NN-512 (**) &  0.006 &  0.001 \\
		\end{tabular}
	\end{table}
	
	Table~\ref{tbl:timeeval} shows the memory and execution time of the proposed approaches in comparison to state-of-the-art methods. The memory consumption was measured with the windows task manager during the execution of each tool. For the runtime, we only measured the execution step of the prediction step. For MMseqs2 we used the total sum of the reported durations of the tool since there was no explicit output for the execution of the easy search itself. Measuring the time consumption in the code of MMseqs2 is difficult since it is optimized for parallelization and execution with special CPU functions like SSE2, SSE4 etc. This was the reason we did not limit the number of threads for MMseqs2, and we allowed it to use the optimal number of threads (32 based on the used CPU). Since BERTax uses pytorch which is only an API-interface to a precompiled C++ code, we decided to implement the bagged decision trees and the neural network in C++, too. This makes their execution much faster and also removes the overhead of Matlab itself. In addition, the tree structure used by Matlab for the bagged decision trees is huge in memory (up to 60 GB after loading) which was removed by the usage of C++ code too. For the subspace KNN, we used the Matlab implementation, which means that 2 GB of the 16 GB are the memory usage of Matlab alone without the loaded subspace KNN. Comparing the runtime of each method, it is obvious that the neural network is the fastest approach, but it does not deliver the best results. It also has the disadvantage of BERTax, that it is not as easy to interpret as the other methods. On the second place in terms of speed are the bagged decision trees. They have a slightly larger memory consumption compared to BERTax, but can be executed much faster. In addition, the decision trees are easier to interpret since the learned thresholds can be directly mapped to the $k$-mer distributions. The third place in terms of runtime goes to the subspace KNN, but this is a deceptive result since the subspace KNN is similar to the MMseqs2 method and needs to store the data set itself for the comparison. Therefore, a larger data set results in an increased runtime. In terms of interpretability, it is as good as the bagged decision trees, since the comparison is done directly on the $k$-mer distribution entries. The only method which is better to interpret is MMseqs2, since it gives the researcher also the local alignments on the nucleotides of the DNA sequence. Taking all results into account (Table~\ref{tbl:sotaeval} and Table~\ref{tbl:timeeval}) the neural network with one hidden layer is the best trade of between MAP of phylum, MAP of superkingdom, memory consumption, and execution time. For the genus level, the best option would be the subspace KNN, but this approach has a larger demand in terms of memory consumption. Therefore, the bagged decision trees, with the 3-mer distribution as feature, would be better if the execution time per sequence and memory consumption are limited.
	
	\section{Discussion and Limitations}
	The length of the sequence is one important factor to select a good $k$-mer size, and different machine learning methods can perform better based on the given data set. While this is a limitation of our approach, it is still easy to apply for every researcher and easy to integrate into taxonomy classification tools like MMseqs2 for example. Another thing that has to be mentioned here too is that the evaluation for DeepMicrobes was performed with a 12-mer from the authors of the BERTax paper. Based on our experience during this research, a 4-mer or 3-mer would be a better choice. Overall, we do not claim to propose the best method in terms of MAP but a simple and easy to use approach for every researcher. In addition, our approach can be used for a pre-selection in the MMseqs2 software before the local alignment is applied, for example. For further research, our approach could also be used as a baseline for large deep neural networks. Another limitation of our evaluation is the data set we have used, since all sequences have an exact length of 1500 nt. For high throughput sequencing technologies, usually millions of 50–200 nt sequences are produced. This can have an effect on the result since different length of sequences have to be evaluated, and they are not cut out of a fixed length sequence like in the used data set~\citep{simon2019benchmarking}. If there are larger sequences available it is also possible, that $k$-mers with a higher k value are more efficient for the classification, and we do not know if the data set introduces a bias into our evaluation. The next limitation is the training of the simple machine learning approaches. Here, a computer with a large amount of memory is necessary due to the implementation of the training procedure in Matlab2022b. In general, it is not required to store all data samples in the memory during training, but it makes the training much faster. This is possibly the root cause why Matlab2022b does load all data samples into the memory. While this limitation is only based on the implementation, it still limits the amount of training data a researcher can use for training or requires using a PC with large amounts of RAM or virtual memory on a flash hard drive. The current implementation of our balancing approach also has a limitation, since we use the ismember function from Matlab. This function requires more computational time compared to a hashing function, but the hashing function would require large amounts of memory, especially for larger $k$-mer distributions or large grid sizes. 
	
	\section{Conclusion}
	In this paper, we propose a $k$-mer distribution as feature for taxonomy classification. This feature together with the proposed feature space data balancing algorithm achieves comparable results to the current state-of-the-art algorithms by using only machine learning based approaches like the bagged decision trees, subspace KNNs or small neural networks. Those approaches are easy to train and do not consume the amount of resources which are necessary for local alignments and comparison as it is used in MMseqs2. Compared to the transformer architecture from BERTax or the deep neural network used in DeepMicropes the simple machine learning algorithm require no specialized hardware, consume less resources, and can be executed very efficiently on any computer or laptop. Based on the data set we showed, that different $k$-mers should be used for optimal performance. Overall, we think the proposed approach is a good baseline for future deep learning approaches in taxonomy classification and an easy-to-use approach for any researcher to classify metagenomic data sets.

	\bibliographystyle{plain}
	\bibliography{template}

\begin{thebibliography}{10}

\bibitem{ainsworth2017k}
David Ainsworth, Michael~JE Sternberg, Come Raczy, and Sarah~A Butcher.
\newblock k-slam: accurate and ultra-fast taxonomic classification and gene
  identification for large metagenomic data sets.
\newblock {\em Nucleic acids research}, 45(4):1649--1656, 2017.

\bibitem{altschul1990basic}
Stephen~F Altschul, Warren Gish, Webb Miller, Eugene~W Myers, and David~J
  Lipman.
\newblock Basic local alignment search tool.
\newblock {\em Journal of molecular biology}, 215(3):403--410, 1990.

\bibitem{brown2016sourmash}
C~Titus Brown and Luiz Irber.
\newblock sourmash: a library for minhash sketching of dna.
\newblock {\em Journal of Open Source Software}, 1(5):27, 2016.

\bibitem{buchfink2015fast}
Benjamin Buchfink, Chao Xie, and Daniel~H Huson.
\newblock Fast and sensitive protein alignment using diamond.
\newblock {\em Nature methods}, 12(1):59--60, 2015.

\bibitem{bui2020cdkam}
Van-Kien Bui and Chaochun Wei.
\newblock Cdkam: a taxonomic classification tool using discriminative k-mers
  and approximate matching strategies.
\newblock {\em BMC bioinformatics}, 21(1):1--13, 2020.

\bibitem{busia2019deep}
Akosua Busia, George~E Dahl, Clara Fannjiang, David~H Alexander, Elizabeth
  Dorfman, Ryan Poplin, Cory~Y McLean, Pi-Chuan Chang, and Mark DePristo.
\newblock A deep learning approach to pattern recognition for short dna
  sequences.
\newblock {\em BioRxiv}, page 353474, 2019.

\bibitem{chiu2019clinical}
Charles~Y Chiu and Steven~A Miller.
\newblock Clinical metagenomics.
\newblock {\em Nature Reviews Genetics}, 20(6):341--355, 2019.

\bibitem{d2016comprehensive}
Rosalinda D’Amore, Umer~Zeeshan Ijaz, Melanie Schirmer, John~G Kenny, Richard
  Gregory, Alistair~C Darby, Migun Shakya, Mircea Podar, Christopher Quince,
  and Neil Hall.
\newblock A comprehensive benchmarking study of protocols and sequencing
  platforms for 16s rrna community profiling.
\newblock {\em BMC genomics}, 17(1):1--20, 2016.

\bibitem{edgar2018updating}
Robert~C Edgar.
\newblock Updating the 97\% identity threshold for 16s ribosomal rna otus.
\newblock {\em Bioinformatics}, 34(14):2371--2375, 2018.

\bibitem{fernandes2021novel}
Marcelo~AC Fernandes and HT~Kung.
\newblock A novel training strategy for deep learning model compression applied
  to viral classifications.
\newblock In {\em 2021 International Joint Conference on Neural Networks
  (IJCNN)}, pages 1--9. IEEE, 2021.

\bibitem{ferragina2000opportunistic}
Paolo Ferragina and Giovanni Manzini.
\newblock Opportunistic data structures with applications.
\newblock In {\em Proceedings 41st annual symposium on foundations of computer
  science}, pages 390--398. IEEE, 2000.

\bibitem{fiannaca2018deep}
Antonino Fiannaca, Laura La~Paglia, Massimo La~Rosa, Lo~Bosco, Giovanni Renda,
  Riccardo Rizzo, Salvatore Gaglio, Alfonso Urso, et~al.
\newblock Deep learning models for bacteria taxonomic classification of
  metagenomic data.
\newblock {\em BMC bioinformatics}, 19(7):61--76, 2018.

\bibitem{jones2001scipy}
Eric Jones, Travis Oliphant, Pearu Peterson, et~al.
\newblock Scipy: Open source scientific tools for python.
\newblock {\em Website}, 2001.

\bibitem{karagoz2021taxonomic}
Meryem~Alt{\i}n Karag{\"o}z and O~Ufuk Nalbantoglu.
\newblock Taxonomic classification of metagenomic sequences from relative
  abundance index profiles using deep learning.
\newblock {\em Biomedical Signal Processing and Control}, 67:102539, 2021.

\bibitem{kim2016centrifuge}
Daehwan Kim, Li~Song, Florian~P Breitwieser, and Steven~L Salzberg.
\newblock Centrifuge: rapid and sensitive classification of metagenomic
  sequences.
\newblock {\em Genome research}, 26(12):1721--1729, 2016.

\bibitem{knights2011bayesian}
Dan Knights, Justin Kuczynski, Emily~S Charlson, Jesse Zaneveld, Michael~C
  Mozer, Ronald~G Collman, Frederic~D Bushman, Rob Knight, and Scott~T Kelley.
\newblock Bayesian community-wide culture-independent microbial source
  tracking.
\newblock {\em Nature methods}, 8(9):761--763, 2011.

\bibitem{li2018minimap2}
Heng Li.
\newblock Minimap2: pairwise alignment for nucleotide sequences.
\newblock {\em Bioinformatics}, 34(18):3094--3100, 2018.

\bibitem{liang2020deepmicrobes}
Qiaoxing Liang, Paul~W Bible, Yu~Liu, Bin Zou, and Lai Wei.
\newblock Deepmicrobes: taxonomic classification for metagenomics with deep
  learning.
\newblock {\em NAR Genomics and Bioinformatics}, 2(1):lqaa009, 2020.

\bibitem{lin2022language}
Zeming Lin, Halil Akin, Roshan Rao, Brian Hie, Zhongkai Zhu, Wenting Lu, Nikita
  Smetanin, Allan dos Santos~Costa, Maryam Fazel-Zarandi, Tom Sercu, Sal
  Candido, et~al.
\newblock Language models of protein sequences at the scale of evolution enable
  accurate structure prediction.
\newblock {\em bioRxiv}, 2022.

\bibitem{loman2013culture}
Nicholas~J Loman, Chrystala Constantinidou, Martin Christner, Holger Rohde,
  Jacqueline Z-M Chan, Joshua Quick, Jacqueline~C Weir, Christopher Quince,
  Geoffrey~P Smith, Jason~R Betley, et~al.
\newblock A culture-independent sequence-based metagenomics approach to the
  investigation of an outbreak of shiga-toxigenic escherichia coli o104: H4.
\newblock {\em Jama}, 309(14):1502--1510, 2013.

\bibitem{louca2019census}
Stilianos Louca, Florent Mazel, Michael Doebeli, and Laura~Wegener Parfrey.
\newblock A census-based estimate of earth's bacterial and archaeal diversity.
\newblock {\em PLoS biology}, 17(2):e3000106, 2019.

\bibitem{louca2021response}
Stilianos Louca, Florent Mazel, Michael Doebeli, and Laura~Wegener Parfrey.
\newblock Response to “vast (but avoidable) underestimation of global
  biodiversity”.
\newblock {\em PLoS biology}, 19(8):e3001362, 2021.

\bibitem{menzel2016fast}
Peter Menzel, Kim~Lee Ng, and Anders Krogh.
\newblock Fast and sensitive taxonomic classification for metagenomics with
  kaiju.
\newblock {\em Nature communications}, 7(1):1--9, 2016.

\bibitem{miller2013metagenomics}
Ruth~R Miller, Vincent Montoya, Jennifer~L Gardy, David~M Patrick, and Patrick
  Tang.
\newblock Metagenomics for pathogen detection in public health.
\newblock {\em Genome medicine}, 5(9):1--14, 2013.

\bibitem{mock2022taxonomic}
Florian Mock, Fleming Kretschmer, Anton Kriese, Sebastian B{\"o}cker, and Manja
  Marz.
\newblock Taxonomic classification of dna sequences beyond sequence similarity
  using deep neural networks.
\newblock {\em Proceedings of the National Academy of Sciences},
  119(35):e2122636119, 2022.

\bibitem{morgulis2008database}
Aleksandr Morgulis, George Coulouris, Yan Raytselis, Thomas~L Madden, Richa
  Agarwala, and Alejandro~A Sch{\"a}ffer.
\newblock Database indexing for production megablast searches.
\newblock {\em Bioinformatics}, 24(16):1757--1764, 2008.

\bibitem{ounit2015clark}
Rachid Ounit, Steve Wanamaker, Timothy~J Close, and Stefano Lonardi.
\newblock Clark: fast and accurate classification of metagenomic and genomic
  sequences using discriminative k-mers.
\newblock {\em BMC genomics}, 16(1):1--13, 2015.

\bibitem{pavia2011viral}
Andrew~T Pavia.
\newblock Viral infections of the lower respiratory tract: old viruses, new
  viruses, and the role of diagnosis.
\newblock {\em Clinical Infectious Diseases}, 52(suppl\_4):S284--S289, 2011.

\bibitem{pedersen2016human}
Helle~Krogh Pedersen, Valborg Gudmundsdottir, Henrik~Bj{\o}rn Nielsen, Tuulia
  Hyotylainen, Trine Nielsen, Benjamin~AH Jensen, Kristoffer Forslund, Falk
  Hildebrand, Edi Prifti, Gwen Falony, et~al.
\newblock Human gut microbes impact host serum metabolome and insulin
  sensitivity.
\newblock {\em Nature}, 535(7612):376--381, 2016.

\bibitem{rojas2019genet}
Mateo Rojas-Carulla, Ilya Tolstikhin, Guillermo Luque, Nicholas Youngblut, Ruth
  Ley, and Bernhard Sch{\"o}lkopf.
\newblock Genet: Deep representations for metagenomics.
\newblock {\em arXiv preprint arXiv:1901.11015}, 2019.

\bibitem{shang2021cheer}
Jiayu Shang and Yanni Sun.
\newblock Cheer: hierarchical taxonomic classification for viral metagenomic
  data via deep learning.
\newblock {\em Methods}, 189:95--103, 2021.

\bibitem{simon2019benchmarking}
H~Ye Simon, Katherine~J Siddle, Daniel~J Park, and Pardis~C Sabeti.
\newblock Benchmarking metagenomics tools for taxonomic classification.
\newblock {\em Cell}, 178(4):779--794, 2019.

\bibitem{somasekar2017viral}
Sneha Somasekar, Deanna Lee, Jody Rule, Samia~N Naccache, Mars Stone, Michael~P
  Busch, Corron Sanders, William~M Lee, and Charles~Y Chiu.
\newblock Viral surveillance in serum samples from patients with acute liver
  failure by metagenomic next-generation sequencing.
\newblock {\em Clinical Infectious Diseases}, 65(9):1477--1485, 2017.

\bibitem{steinegger2017mmseqs2}
Martin Steinegger and Johannes S{\"o}ding.
\newblock Mmseqs2 enables sensitive protein sequence searching for the analysis
  of massive data sets.
\newblock {\em Nature biotechnology}, 35(11):1026--1028, 2017.

\bibitem{truong2015metaphlan2}
Duy~Tin Truong, Eric~A Franzosa, Timothy~L Tickle, Matthias Scholz, George
  Weingart, Edoardo Pasolli, Adrian Tett, Curtis Huttenhower, and Nicola
  Segata.
\newblock Metaphlan2 for enhanced metagenomic taxonomic profiling.
\newblock {\em Nature methods}, 12(10):902--903, 2015.

\bibitem{venkatesan2013case}
Arun Venkatesan, Allan~R Tunkel, Karen~C Bloch, AS~Lauring, J~Sejvar, A~Bitnun,
  JP~Stahl, A~Mailles, M~Drebot, CE~Rupprecht, et~al.
\newblock Case definitions, diagnostic algorithms, and priorities in
  encephalitis: consensus statement of the international encephalitis
  consortium.
\newblock {\em Clinical Infectious Diseases}, 57(8):1114--1128, 2013.

\bibitem{10.1371/journal.pbio.3001192}
John~J. Wiens.
\newblock Vast (but avoidable) underestimation of global biodiversity.
\newblock {\em PLOS Biology}, 19(8):1--4, 08 2021.

\bibitem{wood2019improved}
Derrick~E Wood, Jennifer Lu, and Ben Langmead.
\newblock Improved metagenomic analysis with kraken 2.
\newblock {\em Genome biology}, 20(1):1--13, 2019.

\bibitem{yarza2014uniting}
Pablo Yarza, Pelin Yilmaz, Elmar Pruesse, Frank~Oliver Gl{\"o}ckner, Wolfgang
  Ludwig, Karl-Heinz Schleifer, William~B Whitman, Jean Euz{\'e}by, Rudolf
  Amann, and Ramon Rossell{\'o}-M{\'o}ra.
\newblock Uniting the classification of cultured and uncultured bacteria and
  archaea using 16s rrna gene sequences.
\newblock {\em Nature Reviews Microbiology}, 12(9):635--645, 2014.

\bibitem{zhang2016viral}
Wen Zhang, Linlin Li, Xutao Deng, Johannes Bl{\"u}mel, C~Micha N{\"u}bling,
  Andreas Hunfeld, Sally~A Baylis, and Eric Delwart.
\newblock Viral nucleic acids in human plasma pools.
\newblock {\em Transfusion}, 56(9):2248--2255, 2016.

\end{thebibliography}

\end{document}